\documentclass{PoS}
\usepackage{macros,epsfig,bbm}

\PoS{PoS(LAT2007)084}
\title{Exploring the epsilon regime with twisted mass fermions}

\ShortTitle{Exploring the epsilon regime with twisted mass fermions}

\author{K. Jansen, A. Nube, \speaker{A. Shindler}\thanks{Current address: 
    Theoretical Physics Division, Dept. of Mathematical Sciences,
    University of Liverpool, \newline Liverpool L69 3BX, UK. E-mail:{\tt andrea.shindler@liverpool.ac.uk}}\\
        NIC/DESY Zeuthen, Platanenallee 6 \\
        D-15738 Zeuthen, Germany\\
        E-mail: \email{andrea.shindler@desy.de},\email{karl.jansen@desy.de},\email{annube@ifh.de}}

\author{C. Urbach\\
        Theoretical Physics Division, Dept. of Mathematical Sciences, University of Liverpool \\
        Liverpool L69 7ZL, UK\\
        E-mail: \email{carsten.urbach@liverpool.ac.uk}}

\author{U. Wenger\\
        Institut f\"ur Theoretische Physik, ETH Z\"urich \\
        CH-8093 Z\"urich, Switzerland\\
        E-mail: \email{wenger@itp.phys.ethz.ch}}

\abstract{   
\begin{center}
      \includegraphics[draft=false,scale=1]{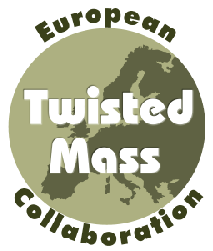}
    \end{center}
  \vskip 0.5cm

In this proceeding contribution we report on a first study in order to explore
the so called \mbox{$\epsilon$ regime} with Wilson twisted mass (Wtm) fermions. 
To show the potential of this approach we 
give a preliminary determination of the chiral condensate.}

\FullConference{The XXV International Symposium on Lattice Field Theory\\
		 July 30 - August 4 2007\\
		 Regensburg, Germany}

\begin{document}

\section{Introduction}
\label{sec:intro}

Simulations of lattice QCD in the so called $\epsilon$ regime~\cite{Gasser:1987ah} of the 
chiral expansion allow in principle the extraction of physical parameters, 
like decay constants and electroweak effective couplings.
Simulations in the $\epsilon$ regime are not exclusive to lattice
actions with exact lattice chiral symmetry. 
While topology certainly plays an important role in this extreme regime, here
we aim at simulations which sample all the topological sectors.
The motivation for this is that the original expressions are given for the
situation where topology is summed over.

To fix the notation we recall that we use 
as the lattice action for $N_{\rm f} = 2$ degenerate flavours
\be
S[\chi,\chibar,U] = S_G[U] + S_F[\chi,\chibar,U] ,
\ee
where $S_G$ denotes the tree-level Symanzik improved gauge action~\cite{Weisz:1982zw}
\be
S_G[U] = \frac{\beta}{3}\sum_x\left\{b_0\sum_{\mu<\nu}
\mathbb{R}{\rm e}~ \Tr \left[ \mathbbm{1} - P^{(1\times 1)}(x;\mu,\nu)\right] + b_1 \sum_{\mu \neq \nu}
\mathbb{R}{\rm e}~ \Tr \left[ \mathbbm{1} - P^{(2\times 1)}(x;\mu,\nu)\right] \right\}
\ee
with the normalization condition $b_0 = 1-8b_1$, and $b_1 = - {1
    \over 12}$.
The fermionic part of the action is given by the Wtm action~\cite{Frezzotti:2000nk,Frezzotti:2003ni}
\be
  S_{\rm F}[\chi,\chibar,U] =a^4\sum_x\chibar(x)\Big[D_{\rm W} + i\mu_{\rm q}\gamma_5\tau^3\Big]\chi(x), 
\label{eq:WtmQCD}
\ee
where 
\be
D_{\rm W} = \frac{1}{2}\{\gamma_\mu(\nabla_\mu + \nabla^*_\mu) -a  \nabla^*_\mu\nabla_\mu\} + m_0,
\label{eq:Wilson}
\ee
$\nabla_\mu$, $\nabla^*_\mu$ are the standard gauge covariant forward and backward
derivatives, $m_0$ and $\mu_{\rm q}$ are respectively 
the bare untwisted and twisted quark masses (for a recent review on Wtm and unexplained notations
see ref.~\cite{Shindler:2007vp}).

The chiral phase diagram of Wilson-like lattice actions, due to lattice artefacts, is different
from the one in the continuum~\cite{Farchioni:2004us,Farchioni:2004fs,Farchioni:2005tu}.
In particular for our choice of the gauge and fermion action, 
the so called first order Sharpe-Singleton scenario~\cite{Sharpe:1998xm} takes place. In this case at 
fixed lattice spacing $a$, if we set $m_{\rm PCAC}=0$ where
\be
m_{\rm PCAC} = \frac{\sum_{\bf x} \langle \partial_0 A_0^a(x) P^a(0)\rangle}
{2\sum_{\bf x} \langle  P^a(x) P^a(0)\rangle} \qquad a=1,2,
\label{eq:PCAC_lat}
\ee
the values of the twisted mass which can be simulated, 
are bounded from below by a value $\mu_{\rm c}$ which is proportional
to $a^2 \Lambda_{\rm QCD}^3$~\cite{Munster:2003ba,Sharpe:2004ps,Scorzato:2004da}.
This result is obtained if one analyzes the potential of the chiral lagrangian 
describing the low energy properties of the theory taking into account also the cutoff effects
of the lattice action up to O($a^2$). In particular this analysis 
is performed in infinite volume.

It is well known~\cite{Gasser:1987ah} that in continuum QCD if one sends 
the quark mass to zero keeping the size of the volume fixed,
chiral symmetry is restored and no phase transition appears,
i.e. the dependence of the chiral condensate on the quark mass is smooth.
It is plausible to expect that at finite lattice spacing the same mechanism takes place,
and in particular the critical point $\mu_{\rm c}$ smooths out.
One way to show this is to include the effects of the non vanishing 
lattice spacing in the analysis of~\cite{Gasser:1987ah}, and to study the 
mass dependence of the chiral condensate in the $\epsilon$ regime including the O($a^2$) effects.
This work is currently in progress.

\section{Algorithm}
\label{sec:algo}

\begin{figure}[h]
\vspace{-1.0cm}
  \begin{center}
    \epsfig{file=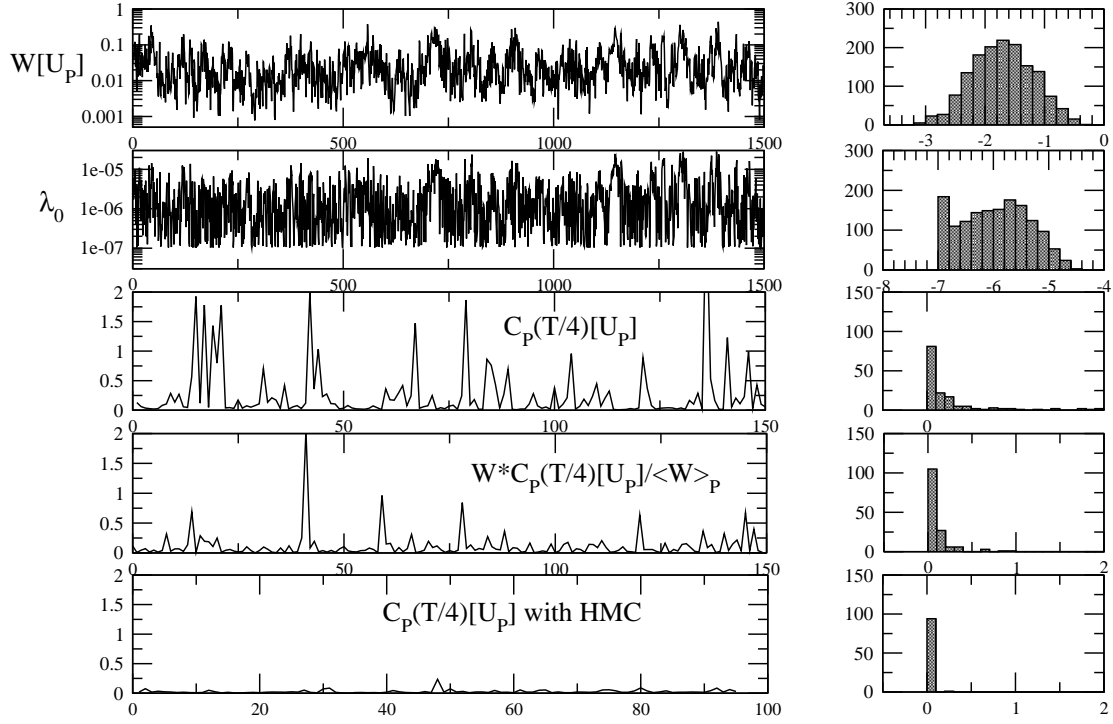,width=13cm,angle=270}
    \vspace{-1.5cm}
    \caption{In the first stripe we plot the MC history and distribution
of the reweighting factor $W$. In the second stripe we plot
the MC history and distribution of the smallest eigenvalue $\lambda_0$ 
of $QQ^\dagger $.
In the remaining three stripes we compare the MC histories and distributions
of the pseudoscalar density correlator at time slice $x_0=T/4$ for the 
PHMC without and with reweighting factor and for the mt-mHMC.}
    \label{fig:eps1}
  \end{center}
\end{figure}

To simulate dynamical Wtm $N_f=2$ degenerate quarks in the $\epsilon$ regime, we
propose to use a PHMC algorithm~\cite{Frezzotti:1997ym,Frezzotti:1998eu,Frezzotti:1998yp,Chiarappa:2005mx}.
We refer to these references for more details on the algorithm, 
and here we just shortly summarize the main ingredients.
The expectation value of a generic observable ${\mathcal O}$ is given by
\be
\langle {\mathcal O} \rangle = \frac{1}{{\mathcal Z}}\int {\mathcal D}[U] {\rm
  e}^{-S_G[U]} \det\left(QQ^\dagger \left[U\right]\right) {\mathcal O}\left[U\right] \quad 
Q = \gamma_5\left[D_{\rm W} + i\mu_{\rm q} \gamma_5\right],
\ee 
with $Q$ being a single flavour operator, which in the following is normalized to have the biggest eigenvalue equal to one.
It can be computed splitting up the determinant in two terms
\be
\det\left(QQ^\dagger \left[U\right]\right)  = \frac{\det\left[QQ^\dagger 
    P_{n,\tilde{\epsilon}}\left(QQ^\dagger \right)\right]}{\det\left[P_{n,\tilde{\epsilon}}\left(QQ^\dagger \right)\right]} , 
\quad P_{n,\tilde{\epsilon}}\left(QQ^\dagger \right) \simeq \left[QQ^\dagger\right]^{-1} ,
\ee
where $P_{n,\tilde{\epsilon}}\left(QQ^\dagger \right)$ is a polynomial approximation of order $n$ for
$\left[QQ^\dagger\right]^{-1}$ restricted in the region of the eigenvalue spectrum $[\tilde{\epsilon},1]$.
The main idea is to split the eigenvalue spectrum of the Wtm operator
into parts which are then treated by either incorporating them in the update
step of a simulation algorithm or by taking them into account in a
reweighting procedure. Following~\cite{Frezzotti:1997ym}, and leaving out the stochastic estimate of the correction
factor, the generic expectation value can be written as 
\be
\langle {\mathcal O} \rangle = \frac{\langle {\mathcal O} W \rangle_P}{\langle
  W \rangle_P}, \quad W =
{\det\left[QQ^\dagger P_{n,\tilde{\epsilon}}\left(QQ^\dagger \right)\right]} \simeq
\prod_{\lambda_i < \tilde{\epsilon}}\left[\lambda_iP_{n,\tilde{\epsilon}}\left(\lambda_i\right)\right] ,
\ee
and the reweighting factor $W$ is computed exactly evaluating $N_\lambda\simeq 20$
smallest eigenvalues of the Wtm operator
where $\lambda_i$ denotes an eigenvalue of $QQ^\dagger $.
This strategy should allow a better sampling of the configuration space, when
the eigenvalues of the Wtm are particularly small.
To be specific in the exploratory runs we have performed we have used a polynomial approximation
with $n=380$ and a cutoff in the spectrum $\tilde{\epsilon} = 5 \cdot 10^{-5}$.
The bare parameters used in the simulations are $\beta=3.9$, $a\mu_{\rm q} = 5\cdot10^{-4}$ and $\kappa = 0.160856$
\cite{Boucaud:2007uk}.
In fig.~\ref{fig:eps1} we summarize first preliminary 
results concerning the quality of the runs. 
In the first stripe we plot the MC history and distribution
of the reweighting factor $W$. We observe a very smooth behaviour and a well behaved distribution.
This has to be compared with the MC history and distribution of the smallest eigenvalue of $QQ^\dagger $, which is
very small, but with fluctuations which are under control.
In the remaining three stripes we compare the MC histories and distributions
of the pseudoscalar density correlator~(\ref{eq:PP}) at time slice $x_0=T/4$ for the 
PHMC without and with reweighting factor and for the mass preconditioned HMC with 
multiple time scale integrator~\cite{Urbach:2005ji}(mt-mHMC).
This plot shows the different sampling of the configuration space by the two algorithms, 
and in particular the HMC seems to suppress contributions coming from 
the very low eigenvalues.These  plots also show the expected advantages of the PHMC algorithm: 
better statistics for the large fluctuations and supression 
of exceptional fluctuations by the exact reweighting factor.

\section{Sampling}
\label{sec:sampl}

\begin{figure}[h]
  \vspace{-0.cm}
\hspace{-1.0cm}
    \epsfig{file=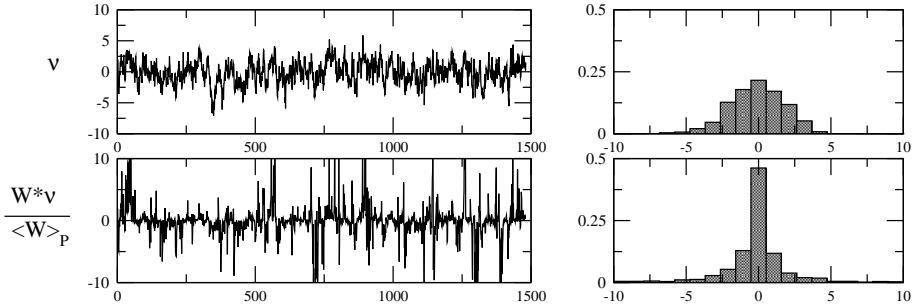,width=11cm,angle=270}
    \vspace{-5.0cm}
    \caption{Comparison of the MC histories and distributions of the topological charge,
      obtained with one level of HYP smearing.
      We observe that the topological charge distribution $N_\nu$, after introducing the reweighting
      factor, has a qualitatively different shape, even if it shows that the algorithm is
      able to sample many topological sectors without any problem.}
    \label{fig:eps2}
\end{figure}

To check that the algorithm correctly samples all the configuration space we have
measured the topological charge $\nu$ using the field theoretical definition 
(proportional to $F\widetilde{F}$) on cooled and APE~\cite{Albanese:1987ds} or HYP~\cite{Hasenfratz:2001hp}
smeared gauge configurations.
We emphasize that the aim of this exercise is just to have a first impression 
on how the algorithm is sampling the configuration space and it is not intended to be an attempt
to compute physical quantities like the topological susceptibility.
\begin{figure}
  \vspace{0.5cm}
  \hspace{3.5cm}
    \epsfig{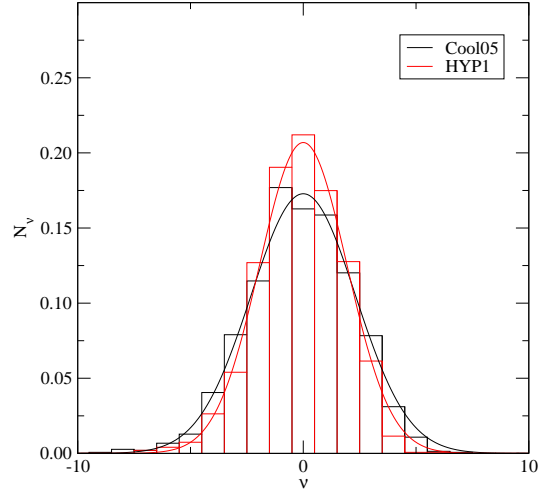}
    \caption{Distribution of the topological charge obtained with the field theoretical
      defintion on cooled and HYP smeared gauge configurations.}
    \label{fig:topo_smear}
\end{figure}
If we consider the sample of gauge configurations generated by the polynomial $P_{n,\tilde{\epsilon}}$ we
obtain distributions of the topological charge as the ones in fig~\ref{fig:topo_smear}.
In fig.~\ref{fig:eps2} we compare the MC histories and distribution of the topological charge
and of the minimal eigenvalue of $QQ^\dagger $ obtained with and without reweighting.
We observe that the topological charge distribution $N_\nu$, after introducing the reweighting
factor, has a qualitatively different shape.

\section{Results}
\label{sec:res}

\begin{figure}
  \begin{center}
    \epsfig{file=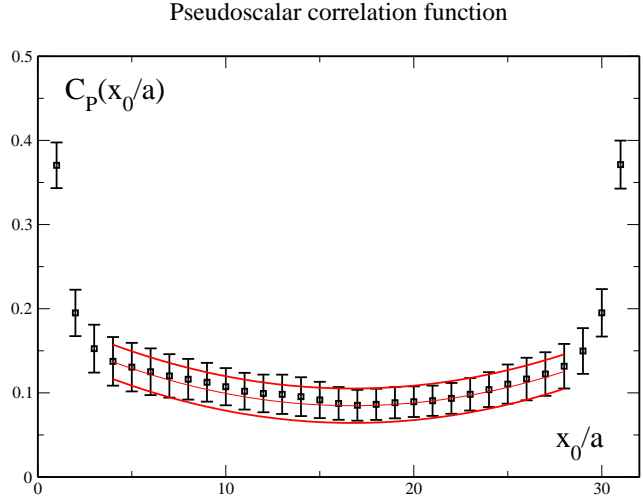,width=8cm, angle=270}
    \caption{Dependence on $x_0/a$ of $C_P(x_0)$ in the $\epsilon$ regime using Wtm fermions.
A parabolic fit and the range used is also shown.}
    \label{fig:pp}
  \end{center}
\end{figure}

A first preliminary result is given by the determination of the 
pseudoscalar density correlator
\be
C_P(x_0) = - a^3 \sum_{\bf x} \langle P^a\left({\bf x},x_0\right)
P^a\left(0\right)\rangle \quad a=1,2 ,
\label{eq:PP}
\ee
where 
\be
P^a\left(x\right) = \chibar(x)\gamma_5\frac{\tau^a}{2}\chi(x) .
\ee
In the $\epsilon$ expansion we have a relation between 
the chiral condensate and $C_P(x_0)$~\cite{Hasenfratz:1989pk,Hansen:1990un}. 
The equivalent relation at fixed topology can be found in~\cite{Damgaard:2001js}.
The correlation function is expected to be constant up to a curvature
in the Euclidean time given by the higher order corrections.
In fig.~\ref{fig:pp} we show the Euclidean time dependence of the correlation 
$C_P(x_0)$, with the fit result and range.
Inserting the values of the lattice spacing and renormalization factors
\be
a = 0.0855(5)(4) , \qquad Z_P^{\overline{MS}}(\mu = 2 GeV) = 0.46(1)(2)
\ee
computed by the ETMC~\cite{Urbach:2007pr,dimopoulos:2007pr,lubicz:2007pr} 
we obtain for the chiral condensate the following determination
\be
|\Sigma_{\rm R}^{\overline{MS}}(\mu = 2 GeV)| = (264(12)(4)^{+20}_{-0} {\rm MeV})^3
\ee
where the first error is the summation in quadrature of all the statistical errors involved
in the determination, the second error is an estimate of the systematic error coming from
the error in the determination of $Z_P$, and the third is an estimate of the higher order corrections
coming from the $\epsilon$ expansion.

This value compares nicely with the $p$ regime determination of ETMC~\cite{Dimopoulos:2007ps}
and of JLQCD~\cite{Fukaya:2007yv}
and other previous determinations summarized in tab.2 of ref.~\cite{McNeile:2005pd}.

\section{Outlooks}
\label{sec:res}

In this proceeding contribution we have shown that with Wtm fermions,
simulations in the $\epsilon$ regime are feasible.
It would be interesting to see if in the current simulations we have a physical volume which  
is big enough to perform a safe matching with chiral perturbation theory.
Work on improving the basic PHMC algorithm we have used so far
is in progress.
Improvements on the quality of the correlation functions could be achieved
with stochastic sources and low mode averaging.
Interesting further investigations
concern, e.g. the distribution of the topological charge, 
and the interplay between the chirally breaking O($a^2$) effects and the contribution of the zero modes
to physical observables. 
Investigations in these directions are in progress.

\section*{Acknowledgments}
We thank the organizers of ``Lattice 2007'' for the very interesting
conference realized in Regensburg. 
A particular acknowledgment goes to Thomas Chiarappa and Roberto Frezzotti
for the work done in preparing a first version of the PHMC code.

\bibliographystyle{JHEP-2}    
\bibliography{epsilon}      

\providecommand{\href}[2]{#2}\begingroup\raggedright\begin{thebibliography}{10}

\bibitem{Gasser:1987ah}
J.~Gasser and H.~Leutwyler {\em Phys. Lett.} {\bf B188} (1987) 477.

\bibitem{Weisz:1982zw}
P.~Weisz {\em Nucl. Phys.} {\bf B212} (1983) 1.

\bibitem{Frezzotti:2000nk}
{\bf ALPHA} Collaboration, R.~Frezzotti, P.~A. Grassi, S.~Sint and P.~Weisz
  {\em JHEP} {\bf 08} (2001) 058
  [\href{http://arXiv.org/abs/hep-lat/0101001}{{\tt hep-lat/0101001}}].

\bibitem{Frezzotti:2003ni}
R.~Frezzotti and G.~C. Rossi {\em JHEP} {\bf 08} (2004) 007
  [\href{http://arXiv.org/abs/hep-lat/0306014}{{\tt hep-lat/0306014}}].

\bibitem{Shindler:2007vp}
A.~Shindler \href{http://arXiv.org/abs/arXiv:0707.4093 [hep-lat]}{{\tt
  arXiv:0707.4093 [hep-lat]}}.

\bibitem{Farchioni:2004us}
F.~Farchioni {\em et.~al.} {\em Eur. Phys. J.} {\bf C39} (2005) 421--433
  [\href{http://arXiv.org/abs/hep-lat/0406039}{{\tt hep-lat/0406039}}].

\bibitem{Farchioni:2004fs}
F.~Farchioni {\em et.~al.} {\em Eur. Phys. J.} {\bf C42} (2005) 73--87
  [\href{http://arXiv.org/abs/hep-lat/0410031}{{\tt hep-lat/0410031}}].

\bibitem{Farchioni:2005tu}
F.~Farchioni {\em et.~al.} {\em Phys. Lett.} {\bf B624} (2005) 324--333
  [\href{http://arXiv.org/abs/hep-lat/0506025}{{\tt hep-lat/0506025}}].

\bibitem{Sharpe:1998xm}
S.~R. Sharpe and J.~Singleton, R. {\em Phys. Rev.} {\bf D58} (1998) 074501
  [\href{http://arXiv.org/abs/hep-lat/9804028}{{\tt hep-lat/9804028}}].

\bibitem{Munster:2003ba}
G.~Munster and C.~Schmidt {\em Europhys. Lett.} {\bf 66} (2004) 652--656
  [\href{http://arXiv.org/abs/hep-lat/0311032}{{\tt hep-lat/0311032}}].

\bibitem{Sharpe:2004ps}
S.~R. Sharpe and J.~M.~S. Wu {\em Phys. Rev.} {\bf D70} (2004) 094029
  [\href{http://arXiv.org/abs/hep-lat/0407025}{{\tt hep-lat/0407025}}].

\bibitem{Scorzato:2004da}
L.~Scorzato {\em Eur. Phys. J.} {\bf C37} (2004) 445--455
  [\href{http://arXiv.org/abs/hep-lat/0407023}{{\tt hep-lat/0407023}}].

\bibitem{Frezzotti:1997ym}
R.~Frezzotti and K.~Jansen {\em Phys. Lett.} {\bf B402} (1997) 328--334
  [\href{http://arXiv.org/abs/hep-lat/9702016}{{\tt hep-lat/9702016}}].

\bibitem{Frezzotti:1998eu}
R.~Frezzotti and K.~Jansen {\em Nucl. Phys.} {\bf B555} (1999) 395--431
  [\href{http://arXiv.org/abs/hep-lat/9808011}{{\tt hep-lat/9808011}}].

\bibitem{Frezzotti:1998yp}
R.~Frezzotti and K.~Jansen {\em Nucl. Phys.} {\bf B555} (1999) 432--453
  [\href{http://arXiv.org/abs/hep-lat/9808038}{{\tt hep-lat/9808038}}].

\bibitem{Chiarappa:2005mx}
T.~Chiarappa, R.~Frezzotti and C.~Urbach
  \href{http://arXiv.org/abs/hep-lat/0509154}{{\tt hep-lat/0509154}}.

\bibitem{Boucaud:2007uk}
{\bf ETM} Collaboration, P.~Boucaud {\em et.~al.} {\em Phys. Lett.} {\bf B650}
  (2007) 304--311 [\href{http://arXiv.org/abs/hep-lat/0701012}{{\tt
  hep-lat/0701012}}].

\bibitem{Urbach:2005ji}
C.~Urbach, K.~Jansen, A.~Shindler and U.~Wenger {\em Comput. Phys. Commun.}
  {\bf 174} (2006) 87--98 [\href{http://arXiv.org/abs/hep-lat/0506011}{{\tt
  hep-lat/0506011}}].

\bibitem{Albanese:1987ds}
{\bf APE} Collaboration, M.~Albanese {\em et.~al.} {\em Phys. Lett.} {\bf B192}
  (1987) 163.

\bibitem{Hasenfratz:2001hp}
A.~Hasenfratz and F.~Knechtli {\em Phys. Rev.} {\bf D64} (2001) 034504
  [\href{http://arXiv.org/abs/hep-lat/0103029}{{\tt hep-lat/0103029}}].

\bibitem{Hasenfratz:1989pk}
P.~Hasenfratz and H.~Leutwyler {\em Nucl. Phys.} {\bf B343} (1990) 241--284.

\bibitem{Hansen:1990un}
F.~C. Hansen {\em Nucl. Phys.} {\bf B345} (1990) 685--708.

\bibitem{Damgaard:2001js}
P.~H. Damgaard, M.~C. Diamantini, P.~Hernandez and K.~Jansen {\em Nucl. Phys.}
  {\bf B629} (2002) 445--478 [\href{http://arXiv.org/abs/hep-lat/0112016}{{\tt
  hep-lat/0112016}}].

\bibitem{Urbach:2007pr}
C.~Urbach {\em PoS} {\bf LAT2007} (2007) 022.

\bibitem{dimopoulos:2007pr}
{\bf ETM} Collaboration, P.~Dimopoulos {\em et.~al.} {\em PoS} {\bf LAT2007}
  (2007) 241.

\bibitem{lubicz:2007pr}
{\bf ETM} Collaboration, V.~Lubicz, S.~Simula and C.~Tarantino {\em PoS} {\bf
  LAT2007} (2007) 374.

\bibitem{Dimopoulos:2007ps}
{\bf ETM} Collaboration, P.~Dimopoulos {\em et.~al.} {\em PoS} {\bf LAT2007}
  (2007) 102.

\bibitem{Fukaya:2007yv}
H.~Fukaya {\em et.~al.} \href{http://arXiv.org/abs/arXiv:0705.3322
  [hep-lat]}{{\tt arXiv:0705.3322 [hep-lat]}}.

\bibitem{McNeile:2005pd}
C.~McNeile {\em Phys. Lett.} {\bf B619} (2005) 124--128
  [\href{http://arXiv.org/abs/hep-lat/0504006}{{\tt hep-lat/0504006}}].

\end{thebibliography}\endgroup

\end{document}